**Effect of protein binding on exposure of unbound and total mycophenolic acid: a population pharmacokinetic analysis in Chinese adult kidney transplant recipients**

## Authors list

Changcheng Sheng[1], Qun Zhao[1], Wanjie Niu[1], Xiaoyan Qiu[1], Ming Zhang[2], Zheng Jiao[1*]

## Authors' institutional affiliations

1 Department of Pharmacy, Huashan Hospital, Fudan University, Shanghai 200040, China

2 Department of Nephropathy, Huashan Hospital, Fudan University, Shanghai, 200040, China

## *Corresponding author

Zheng Jiao, Ph.D

Department of Pharmacy, Huashan Hospital, Fudan University

12 Middle Urumqi Road, Shanghai, 200040, China

Tel: +86 021-5288 8712

Fax: +86 021-6248 6927

Email: zjiao@fudan.edu.cn

## ORCID information

Zheng Jiao: 0000-0001-7999-7162

Chang-Cheng Sheng: 0000-0001-5844-0154

## Abstract

***AIMS*** A population pharmacokinetic (PK) analysis was performed to: (1) characterise the PK of unbound and total mycophenolic acid (MPA) and its 7-O-mycophenolic acid glucuronide (MPAG) metabolite, and (2) identify the clinically significant covariates that cause variability in the dose-exposure relationship to facilitate dose optimisation.

***METHODS*** A total of 740 unbound MPA (uMPA), 741 total MPA (tMPA) and 734 total MPAG (tMPAG) concentration-time data from 58 Chinese kidney transplant patients were analysed using a nonlinear mixed-effect model. The influence of covariates was tested using a stepwise procedure.

***RESULTS*** The PK of unbound MPA and MPAG were characterised by a two- and one-compartment model with first-order elimination, respectively. Apparent clearance of uMPA ($CL_{uMPA}/F$) was estimated to be 852 L/h with a relative standard error (RSE) of 7.1%. The tMPA and uMPA were connected using a linear protein binding model, in which the protein binding rate constant ($k_B$) increased non-linearly with the serum albumin (ALB) concentration. The estimated $k_B$ was 53.4 /h (RSE, 2.3%) for patients with ALB of 40 g/L. In addition, model-based simulation showed that changes in ALB substantially affected tMPA but not uMPA exposure.

***CONCLUSIONS*** The established model adequately described the population PK characteristics of the uMPA, tMPA, and MPAG. The estimated $CL_{uMPA}/F$ and unbound fraction of MPA ($FU_{MPA}$) in Chinese kidney transplant recipients were comparable to those published previously in Caucasians. We recommend monitoring uMPA instead of tMPA to optimise mycophenolate mofetil (MMF) dosing for patients with lower ALB levels.

## Introduction

Mycophenolate mofetil (MMF), a pro-drug of mycophenolic acid (MPA), is the predominant antimetabolite immunosuppressant used as a co-therapy with tacrolimus (TAC) or ciclosporin (CsA) to prevent graft rejection after solid organ transplantation [1, 2]. MMF is extensively absorbed and rapidly hydrolysed to the active component MPA after oral administration. The latter is primarily metabolised to the major but inactive 7-O-mycophenolic acid glucuronide (MPAG) and the relatively minor but active acyl-glucuronide mycophenolic acid (AcMPAG) [3, 4]. MPA and MPAG are reported to be 97% and 82% bound to serum albumin (ALB) at clinically relevant concentrations, respectively [3]. MPAG also undergoes the enterohepatic circulation (EHC) through biliary excretion, intestinal de-glucuronidation by microflora, and reabsorption as MPA in the colon. This process contributes to approximately 40% (range: 10–60%) of the area under the concentration-time curve (AUC) of MPA and causes multiple peaks in the concentration-time profile [3, 5]. Most absorbed MMF is eliminated through the kidney as MPAG [5].

MPA has a narrow therapeutic window. It is recommended to maintain a 12-h dosing interval exposure ($AUC_{0\text{-}12h}$) between 30 and 60 mg ·h/L during the early post-transplantation period when MMF is prescribed with CSA [6-9]. Under-exposure is associated with an increased risk for acute rejection, whereas a higher $AUC_{0\text{-}12h}$ may lead to over-immunosuppression. Large between-subject variability (BSV) and time-dependent variation within-subject are characteristics of MPA pharmacokinetics (PK) [10-12]. A 10-fold variation of MPA exposure was observed even in subjects administered the same dose during the first 2 weeks following kidney transplantation. Moreover, the AUC of MPA in the first few weeks post-transplantation was 30–50% lower than that in the later period following administration of the same dose of MMF [10].

The narrow therapeutic window and large PK variability make it necessary to individualise MMF therapy based on therapeutic drug monitoring (TDM). Currently, the maximum a *posterior* Bayesian (MAPB) method using population PK in combination with Bayesian estimation is recommended for facilitating the optimal pharmacotherapy [13-16]. This approach is based on a comprehensive understanding of *prior* information, i.e. the population PK characteristics.

The population PK characteristics of total MPA (tMPA) in kidney transplant recipients have been extensively investigated in various populations and no obvious ethnic difference has been observed in tMPA population PK [17-26]. There are few investigations of the population PK characteristics of unbound MPA (uMPA) [27-30], the pharmacologically active component, due to the technical complexity of measurements. Furthermore, little is known in Chinese kidney transplant recipients. Therefore, the aims of this study were to develop a population PK model to: (1) characterise the PK of uMPA, tMPA, and the metabolite MPAG, and (2) identify the clinically significant covariates that cause variability in the dose-exposure relationship to facilitate dose optimisation.

## Methods

### Study design and patients

The data were obtained from two clinical studies conducted in 58 Chinese adult kidney transplant recipients [31, 32]. All recipients received triple immunosuppressive therapy comprising MMF (CellCept®, Roche Pharma Ltd., Shanghai, China), CsA, and corticosteroids. The first study was an evaluation of the PK of MPA and MPAG during the early post-transplantation period conducted at Huashan Hospital, Fudan University [31]. MMF was initiated at 1500 mg/day from the day of the surgery. The second study was an open-label, multi-centre, two-phase, sequential, bioequivalence study conducted in stable kidney transplantation patients [32]. MMF dose was 1000 or 1500 mg/day in most patients. All protocols were approved by the independent Clinical Research Ethics Committee of Huashan Hospital, Fudan University and all participants provided written informed consent before enrolment.

After the morning dose, whole blood samples were collected at 0, 0.5, 1, 1.5, 2, 3, 4, 6, 8, 10, and 12 h in study 1, and at 0, 0.5, 1, 1.5, 2, 2.5, 3, 4, 6, 9, 10, and 12 h in study 2. Low-fat meals were provided after the scheduled 4 and 10 h samplings in study 1, and the 3 and 9 h samplings in study 2. The relevant data were collected to explore the relationships between demographic characteristics, biochemical measurements, and PK parameters.

All samples were analysed at Huashan Hospital using a validated high-performance liquid chromatography method [31, 33]. The calibration ranges were 0.002–1.0, 0.1–40, and 10–200 mg/L for uMPA, tMPA, and total MPAG (tMPAG), respectively. The relative bias values were within ±17%, ±8.3%, and ±5.2% for uMPA, tMPA, and tMPAG, respectively. The intra- and inter-day precision, as coefficient of variation values, were < 10% and 14% for uMPA, 5.1% and 9.2% for tMPA, and 6.7% and 9.8% for tMPAG, respectively.

## Population PK analyses

### *Software and model selection criteria*

Population PK analyses were performed using the nonlinear mixed-effect modelling software (NONMEM®, version 7.4; ICON Development Solutions, Ellicott City, MD, USA) compiled with gfortran 4.6.0. Perl-speaks-NONMEM (PsN, version 4.7.0; http://uupharmacometrics.github.io/PsN) and Pirana (version 2.9.7; http://www.certara.com/pirana) were used to link NONMEM, model development, and model evaluation. The stochastic approximation expectation maximisation (SAEM), followed by important sampling (IMP) method [34] were used throughout the model development. Graphical diagnostics were performed using R software (version 3.4.4, http://www.r-project.org).

MMF doses and uMPA, tMPA, and tMPAG concentrations were transformed into molar equivalents by dividing them with the molecular weight (MMF, MPA, and MPAG: 433.498, 320.339, and 496.462 g/mol, respectively; http://chem.nlm.nih.gov/chemidplus/) and then reconverted to milligram per litre in the figures and results.

Model selection was based on goodness-of-fit (GOF) plots [35] in addition to the three commonly used criteria of statistical significance, plausibility, and stability. The difference in objective function values (OFV) between two nested models was used for statistical comparison. Akaike information criteria (AIC) [36] and Bayesian information criteria (BIC) [37] were used to discriminate non-nested models.

Additionally, relative standard errors (RSEs) of parameter estimates, shrinkages, and changes of BSV and residual unexplained variability (RUV) estimates were considered. During the model developing process, the condition numbers were calculated and no more than 1000 were kept to avoid over-parameterisation [38].

### *Model development*

Population PK modelling of MPA and MPAG was conducted using a sequential approach and eventually led to a simultaneous modelling of both the parent compound and metabolite. One- or two-compartments models with first-order elimination were tested for uMPA and unbound MPAG (uMPAG). We further investigated whether MPA absorption was best described by a first- or zero-order process, with or without a lagged absorption time (Tlag). The concentrations of uMPAG were not determined in our study but were estimated from tMPAG by multiplying the unbound fraction of MPAG ($FU_{MPAG}$), which was fixed at 18% according to a previous study [3].

The tMPA data was first modelled by adding a linear protein binding compartment as equation 1.

$$C_{tMPA} = C_{uMPA} + k_B \times C_{uMPA} \quad (1)$$

where, $C_{tMPA}$ and $C_{uMPA}$ represent total and unbound MPA concentrations, respectively, and $k_B$ is the protein binding rate constant. In this case, the unbound fraction of MPA ($FU_{MPA}$) could be expressed as equation 2.

$$FU_{MPA} = \frac{C_{uMPA}}{C_{tMPA}} = \frac{1}{1+k_B} \quad (2)$$

The non-linear saturable protein binding model published previously [29, 39] was also evaluated using equation 3.

$$C_{bMPA} = \frac{B_{max} \times C_{uMPA}}{k_D + C_{uMPA}} \quad (3)$$

where, $C_{bMPA}$ represents the bound MPA concentration, $B_{max}$ is the maximal number of protein binding sites, and $k_D$ is the dissociation constant representing the uMPA concentration corresponding to half-saturation of protein binding.

To describe the physiological EHC process, the previously published intermittent EHC model [40, 41] was used with some modifications, in which a gallbladder compartment was introduced to connect MPAG and gut compartments. The percentage of MPAG recycled into the systemic circulation (%EHC) was described using equation 4.

$$\%EHC = \frac{k_{GG}}{k_{GG} + k_{e0}} \times 100 \quad (4)$$

where, $k_{GG}$ is the transfer rate constant from the MPAG central compartment to the gallbladder and $k_{e0}$ is the elimination rate constant of MPAG.

Several assumptions were made to ensure the model was structurally identifiable [40]: (1) MMF is completely and quickly absorbed, (2) the conversion ratio from MMF to MPAG is fixed at 100%, (3) MPAG secreted from the gallbladder to intestines is completely deconjugated to MPA and reabsorbed, (4) the rate constants associated with each compartment are all first-order and unaffected by the recycling, and (5) gallbladder emptying is triggered by meals. Additionally, the gallbladder emptying rate constant ($k_{GB}$) was fixed at 3.708 /h based on previous study [42]. The duration ($D_{GB}$) of gallbladder release was fixed at 0.5 h to ensure that over 90% gallbladder contents would be released after each trigger.

An exponential model was used to describe BSVs for each PK parameter. Exponential, additive,

and combined models were compared to describe RUVs. Furthermore, the covariance of BSVs was estimated with OMEGA BLOCK statement in NONME. Since uMPA, tMPA, and tMPAG concentrations were derived from the same sample for each subject, correlations of their RUVs were likely to exist. An L2 data item was introduced and the covariance of RUVs was also evaluated with SIGMA BLOCK statement [34].

After the base model was determined, the following physiologically meaningful covariates were investigated: sex, age, body weight (BW), postoperative time (POT), haemoglobin (HB), ALB, alanine aminotransferase (ALT), aspartate aminotransferase (AST), serum creatinine (SCr), glomerular filtration rate (GFR), CsA daily dose, and co-administration of antacids. GFR was estimated from SCr using the Chronic Kidney Disease Epidemiology Collaboration (CKD-EPI) formula [43].

First, relationships between individual PK parameters and covariates were examined by graphical inspection to identify the potential covariates. Then, the identified covariates were tested using a stepwise procedure. During the forward inclusion and backward elimination steps, significance levels were set at a decrease in OFV > 3.84 ($\chi^2$, $df = 1$, $p < 0.05$) and an increase in OFV > 10.83 ($\chi^2$, $df = 1$, $p < 0.001$), respectively. The continuous covariates were assessed using a linear and non-linear model, and categorical covariates were modelled proportionally. To demonstrate clinical significance, covariates were only retained if the effect on the corresponding parameter was > 15% for a categorical covariate, or > 15% at the highest or lowest observed covariate value for a continuous covariate [44]. In addition, the included covariates were expected to have interpretations of physiological or pharmacological mechanisms.

***Model evaluation***

The established model was evaluated by graphical diagnosis. GOF plots included scatterplots of population predictions (PRED) and individual predictions (IPRED) versus observed concentrations (OBS), as well as conditional weighted residuals (CWRES) versus PRED and time after previous dose (TAD). Additionally, 500 bootstraps [45] were applied to assess the reliability and stability of the final model. The medians and 2.5%–97.5% intervals from the bootstrap replicates were compared with estimates of the final model.

The final model was further checked by a prediction-corrected visual predictive check (pc-VPC) [46] and posterior predictive check (PPC) [47]. Furthermore, 2000 datasets were simulated using the final model from the original dataset. For pc-VPC, the observed and simulated concentrations were dose-normalised to 750 mg MMF every 12 h. The median simulated concentrations and corresponding 5%–95% interval were calculated and graphically compared with the observations. PPC was further performed to assess if the model appropriately predicted the $AUC_{0\text{-}12h}$ of uMPA, tMPA, and tMPAG. Simulated and observed $AUC_{0\text{-}12h}$ were calculated using the linear trapezoidal rule. Distributions of the simulated and observed $AUC_{0\text{-}12h}$ were then graphically compared.

### Simulation analyses of effects of significant covariates

The established final model was used to investigate the effect of the identified covariates on the PK of MPA and MPAG. Specifically, 2000 stochastic simulations were performed for virtual subjects administered 750 mg MMF every 12 h with different covariate levels. The $AUC_{0\text{-}12h}$ values of uMPA, tMPA, and tMPAG were estimated using the linear trapezoidal rule and changes in $AUC_{0\text{-}12h}$ and $FU_{MPA}$ were assessed.

## Results

### Patient characteristics and data descriptions

A total of 27 full concentration-time profiles containing uMPA, tMPA, and tMPAG data were obtained from 20 patients in study 1, including 23 profiles collected within 3 months post-transplantation. Sixteen patients had one profile, one had two profiles, and the other three had three profiles. In study 2, 38 full concentration-time profiles were obtained from 38 patients, including 37 collected beyond 3 months post-transplantation. The patient characteristics are shown in **Table 1**. Of these subjects, male patients accounted for approximately 78%. The concomitant antacids in study 1 were proton pump inhibitors, whereas sodium hydrogen carbonate and compound aluminium hydroxide were co-administered in study 2. Significant differences in BW, POT, HB, and ALB as well as doses of MMF, CsA, and corticosteroids were observed between the two studies.

Of the 2229 samples, < 1% (3 uMPA, 1 tMPA, and 10 tMPAG) were below the lower limit of quantification and were discarded. In total, 740 uMPA, 741 tMPA, and 734 tMPAG concentration measurements were used for the population PK analysis. Multiple uMPA and tMPA peaks attributed to EHC were observed at 4–6 and 8–12 h post-dosing in some subjects, whereas no obvious multiple peaks were observed for tMPAG.

## Population PK model

### *Model development*

A five-compartment model with first-order absorption and elimination adequately described the uMPA, tMPA, and uMPAG data. The schematic representation of the final structural model is shown in **Figure 1**. The PK profiles of uMPA and uMPAG were characterised using a two- and one-compartment model, respectively. Incorporation of Tlag further led to a significant reduction of 387.232 units in the OFV. Simultaneously estimation of both $B_{max}$ and $k_D$ was not feasible; therefore, $B_{max}$ was fixed at the reported value of 35100 μmol [27]. The non-linear saturable binding from the central compartment did not improve the fit (AIC, 1533.614 *vs.* 1529.537; BIC, 2012.667 *vs.* 2008.59) more than the linear protein binding model did.

The PK parameters estimated were absorption rate constant ($k_a$), Tlag, apparent clearance of uMPA and uMPAG ($CL_{uMPA}$/F and $CL_{uMPAG}$/F, respectively), inter-compartmental clearance of uMPA ($Q_{uMPA}$/F), apparent central volume of distribution of uMPA and uMPAG ($V_{CuMPA}$/F and $V_{CuMPAG}$/F, respectively), $k_B$ and %EHC. Apparent peripheral volume of distribution of uMPA ($V_{PuMPA}$/F) could not be estimated appropriately and was fixed at the reported literature value of 34300 L [27].

Considering that the expectation maximisation (EM) algorithm is much more robust and adept at handling the large full OMEGA block [34], we initially attempted to assign BSVs to all PK parameters. However, our data did not support the estimation of BSV on $k_B$. To maximally enhance the EM efficiency, the BSV was assigned to $k_B$ and its variance was fixed at 0.01 [34]. Various RUV models were tested to describe the residual errors. Incorporation of the additive residual error resulted in boundary issues and, therefore, an exponential RUV model was used.

Based on the visual inspections and clinical plausibility, the effects of BW and sex on $CL_{uMPA}$/F, $Q_{uMPA}$/F, $V_{CuMPA}$/F, $CL_{uMPAG}$/F, and $V_{CuMPAG}$/F; GFR on $k_B$, $CL_{uMPA}$/F, $Q_{uMPA}$/F, and $CL_{uMPAG}$/F; ALB on $k_B$, $V_{CuMPA}$/F, and $V_{CuMPAG}$/F; co-administration of antacids on $CL_{uMPA}$/F and $k_a$; and tMPAG concentrations on $k_B$ were further tested using the stepwise method. Of these, the effects of ALB on $k_B$, GFR on $CL_{uMPAG}$/F, BW and sex on $Q_{uMPA}$/F, and co-administration of antacids on $k_a$ were included in the forward procedure, whereas the effects of co-administration of antacids on $k_a$ and sex on $Q_{uMPA}$/F showed no significance in the backward step and, thus, were not retained in the final model. The forward inclusion and backward elimination steps are summarised in **Table S1**.

Although introduction of full variance-covariance matrices for BSVs and RUVs substantially decreased the OFV by 288.048 units, the high condition number ($1.94 \times 10^{10}$) indicated that the model might be ill-conditioned because of over-parameterisation. Finally, the covariances between BSVs for $V_{CuMPAG}$/F and $CL_{uMPAG}$/F, and between RUVs for uMPA and tMPA were included. This further decreased the OFV by 159.478 units with an acceptable condition number (< 150).

The parameter estimates of final model are provided in **Table 2**. No significant covariate was detected to influence $CL_{uMPA}$/F, whereas non-linear relationships were found between $k_B$ and ALB, and between $CL_{uMPAG}$/F and GFR. RSEs of the parameter estimates were < 25% and 45% for fixed and random effects, respectively. Shrinkage values of BSVs and RUVs were < 30% except for %EHC.

### *Model evaluation*

The basic GOF plots of the final model are shown in **Figure 2** where PRED and IPRED did not show obvious bias when plotted against OBS. No evident trend in plots of CWRES versus PRED or TAD were found. Over 95% observations were within ±2 CWRES, indicating that the model was appropriately unbiased and adequately described the data.

Out of 500 replicates in the bootstrap analysis, 495 runs converged successfully. The estimated parameters based on original dataset were in good agreement with the median bootstrap replicates and were within the 2.5%–97.5% intervals obtained from the bootstrap analysis (**Table 2**), indicating the reliability and stability of final model.

**Figure 3** shows the results of the pc-VPC of the final model. Most observed concentrations fell within the 90% prediction interval and no obvious discrepancy between observations and simulations was found. The PPC suggested that the simulated $AUC_{0\text{-}12h}$ values also showed good consistency with the observations (**Figure 4**). The pc-VPC and PPC results showed that the final model was reasonably good at predicting the observations.

### Simulations illustrating effect of covariates

Typical subjects administered 750 mg MMF every 12 h were simulated with different ALB and GFR levels. ALB values were set from 20 to 50 g/L with a step of 5 g/L. At each ALB level, the GFR was set at 15, 30, 60, 90, and 120 mL/min according to the Kidney Disease: Improving Global Outcomes chronic kidney disease classification [48]. Generally, ALB and GFR showed large effects on tMPA and tMPAG, respectively, but little effect on uMPA (**Figure 5**).

A substantial decrease in tMPA $AUC_{0\text{-}12h}$ and increase in $FU_{MPA}$ were observed with decreasing ALB concentrations. For subjects with a GFR of 90 mL/min administered 750 mg MMF every 12 h, the median tMPA $AUC_{0\text{-}12h}$ decreased from 43.03 to 20.66 mg ·h/L when ALB concentrations decreased from 40 to 20 g/L, whereas the exposure of uMPA and tMPAG remained almost unchanged (<5%). A decrease in ALB concentrations from 40 to 20 g/L increased the median $FU_{MPA}$ from 1.86 to 3.80% (**Figure S1**).

Additionally, a substantial increase in tMPAG $AUC_{0\text{-}12h}$ was observed with decreasing GFR. For subjects with an ALB concentration of 40 g/L administered 750 mg MMF every 12 h, a reduction in GFR from 90 to 15 mL/min led to a 3.65-fold increase in median tMPAG $AUC_{0\text{-}12h}$, while $FU_{MPA}$ and the exposure of both uMPA and tMPA were unchanged (**Figures 5 and S1**).

The simulations showed that neither ALB nor GFR significantly affected uMPA exposure. For patients with lower ALB levels, dose adjustment based on monitoring tMPA would lead to higher risk of leucopoenia and infections, resulted by overexposure to uMPA.

## Discussion

The present study extensively investigated the population PK characteristics of uMPA, tMPA, and MPAG in Chinese adult kidney transplant recipients during both the early and stable periods post-transplantation. A two-compartment model with first-order absorption and elimination adequately described the uMPA data. Furthermore, the uMPA and tMPA were connected using a linear protein binding model.

As shown in **Table 3**, the population estimate of $CL_{uMPA}/F$ (852 L/h) was comparable to most of previously reported values in Caucasians (654–866 L/h) [27-29]. Moreover, the calculated typical value of apparent clearance of tMPA ($CL_{tMPA}/F$, 15.68 L/h) was also comparable to most of previously reported values (12.3–20.8 L/h) [17-26]. The population estimate of $FU_{MPA}$ in our study (1.84%) was similar to values reported by van Hest *et al.* [28] and Colom *et al.* [29] (2.03% and 1.93%, respectively). There were no obvious ethnic differences among different populations.

The stepwise covariate analyses suggested that ALB had significant effects on $k_B$ and $FU_{MPA}$. MPA is extensively bound to human ALB, which has more than one binding site on each molecule with equivalent binding characteristics [49]. A reduction in ALB decreases the binding sites, which increases the $FU_{MPA}$.

The simulations showed that changes in ALB concentrations had substantial effects on $FU_{MPA}$ and tMPA exposure, but little effect on uMPA exposure, which was consistent with the findings of de Winter *et al*. [27] and van Hest *et al*. [28]. This could be attributed to the low hepatic extraction ratio of MPA. $FU_{MPA}$ tended to increase with decreasing ALB (**Figure S1**) and the increase in $FU_{MPA}$ caused relatively more uMPA to be metabolised and eliminated from the body, thereby decreasing tMPA exposure. In contrast, MPA is characterised by a low hepatic extraction ratio of 0.2 [50], and the unbound exposure of drugs with low extraction ratio is unaffected by changes in the unbound fraction [51].

These results suggested that dose adjustment based on tMPA exposure might not be appropriate under lower ALB conditions. A reduction in median tMPA $AUC_{0\text{-}12h}$ from approximately 43 to 21 mg ·h/L was observed when ALB concentration decreased from 40 to 20 g/L for patients administered 750 mg MMF every 12 h. However, this observation does not indicate an MMF dose increment is necessary because of the unchanged uMPA exposure. Although relationships between uMPA exposure and the acute rejection risk have not been fully identified, uMPA has been acknowledged as the pharmacologically active component. Moreover, uMPA exposure has been demonstrated to be associated with the risk of leucopoenia and infections [52-55]. An increased MMF dose would also increase uMPA exposure, placing patients at a higher risk of over-immunosuppression with manifestations such as leucopoenia and infections. In such situations, monitoring uMPA exposure might be preferable to monitoring tMPA for adjusting the MMF dose.

Additionally, a gallbladder compartment was introduced to characterise the intermittent EHC process in the present study. The EHC process is mediated by multidrug resistance-associated protein 2 (MRP2), which is inhibited by CsA [56]. All subjects in our study were co-treated with CsA, and concentration-time profiles only showed slight multiple peaks attributed to EHC in some individuals. Therefore, inhibition of EHC by CsA might likely explain why the final model estimated an extremely low %EHC with a high shrinkage (>50%).

Regarding the metabolite MPAG primarily eliminated through the kidneys, statistically significant relationship was found between $CL_{uMPAG}/F$ and kidney function, which was consistent with previous studies [20, 25, 27, 28]. Nevertheless, the previously reported competitive protein binding relationship between MPA and MPAG [27, 28] was not observed, which might be associated with the relatively lower MPAG concentrations (median, 49.79 mg/L). Only 7.6% (56/734) of the tMPAG concentrations were > 100 mg/L with a maximum of 177.9 mg/L in our study. At high concentrations, MPAG could displace MPA from its protein binding sites. It has been reported *in vitro* that $FU_{MPA}$ increased 3-fold as the MPAG concentration increased from 0 to 800 mg/L [49].

There are some limitations in the present study. Firstly, uMPAG data were not available and $FU_{MPAG}$ was fixed to previously reported literature values. Thus, the competitive protein binding process with a mass balance published by de Winter *et al.* [27] could not be further investigated. Secondly, only one dose level of MMF was administered to most patients in our study, which prevented us from investigating the non-linear relationship between MMF dose and MPA exposure as reported by de Winter *et al.* [23]. Lastly, all patients in our study were co-administered with MMF and CsA, therefore, our results might only be applicable to patients co-treated with CsA.

In summary, the established model adequately described the population PK characteristics of uMPA, tMPA, and MPAG. Large BSVs and RUVs were still observed, suggesting TDM would be necessary for optimisation of MMF therapy. The estimated $CL_{uMPA}/F$ and $FU_{MPA}$ in Chinese kidney transplant recipients were comparable to those published previously in Caucasians. In addition, tMPA exposure reduced with decreasing ALB, which had little effect on uMPA exposure. Therefore, under lower ALB conditions, dose adjustment based on tMPA exposure might place patients at higher risk of over-immunosuppression. We recommend monitoring uMPA instead of tMPA to optimise MMF dosing for patients with lower ALB concentrations.

Transplant 2005; 5: 987-94.

**Table 1 Patient characteristics and clinical covariates**

| Characteristic | Study 1 | | Study 2 | | *P* value [a] |
|---|---|---|---|---|---|
| | median (range) | mean ±SD | median (range) | mean ±SD | |
| Patients, n | 20 | / | 38 | / | / |
| Sex | | | | | |
| Male, n (%) | 11 (55) | / | 34 (89) | / | < 0.01 |
| Female, n (%) | 9 (45) | / | 4 (11) | / | < 0.01 |
| Age, years | 36 (19-61) | 37 ±12 | 38 (18-62) | 38 ±12 | > 0.05 |
| Body weight, kg | 55 (40-71) | 54.3 ±9.8 | 65 (42-82.5) | 65.2 ±10.2 | < 0.001 |
| Postoperative time, days | 10 (3-148) | 31 ±41 | 298 (70-3084) | 620 ±780 | < 0.001 |
| Mycophenolate mofetil daily dose, mg/day | 1500 (750-2000) | 1444 ±313 | 1000 (1000-2000) | 1230 ±269 | < 0.01 |
| Hemoglobin, g/L | 86 (72-134) | 93.6 ±18.6 | 139 (103-181) | 142.6 ±22.4 | < 0.001 |
| Albumin, g/L | 31 (20-43) | 32 ±6.6 | 44.9 (32.3-50) | 44.2 ±3.9 | < 0.001 |
| Alanine aminotransferase, U/L | 24 (10-390) | 49.48 ±78.51 | 18 (7-64) | 21.88 ±12.65 | > 0.05 |
| Aspartate aminotransferase, U/L | 20 (7-139) | 33.78 ±29.32 | 24 (8.6-86) | 28.94 ±19.88 | > 0.05 |
| Serum creatinine, μmol/L | 96 (50-443) | 114.41 ±73.97 | 104.5 (76-152.9) | 108.82 ±17.27 | > 0.05 |
| Glomerular filtration rate [b], mL/min | 76.12 (11.17-123.8) | 75.58 ±25.09 | 74.42 (45.14-102.3) | 74.79 ±14.16 | > 0.05 |
| **Concomitant medication** | | | | | |
| Ciclosporin daily dose, mg/day | 300 (0-400) | 282 ±102 | 220 (100-400) | 231 ±65 | < 0.01 |
| Corticosteroid daily dose, mg/day | 20 (5-675) | 49.1 ±126.1 | 10 (3-20) | 10.8 ±4.1 | < 0.001 |
| Antacids [c], n (%) | 6 (22) | / | 5 (13) | / | > 0.05 |
| Aspirin, n (%) | 0 (0) | / | 6 (16) | / | < 0.05 |
| Nifedipine, n (%) | 4 (15) | / | 5 (13) | / | > 0.05 |
| Diltiazem, n (%) | 0 (0) | / | 7 (18) | / | < 0.05 |

/, not applicable; SD, standard deviation

[a] Differences between groups are determined using the Mann-Whitney U test for continuous variables and Fisher's exact test for categorical data with IBM SPSS Statistics for Windows (Version 20, IBM Corp., Armonk, NY).

[b] Glomerular filtration rate (GFR) is calculated from serum creatinine using the Chronic Kidney Disease Epidemiology Collaboration (CKD-EPI) formula [43]: $GFR = 141 \times \min(SCr/\kappa, 1)^{\alpha} \times \max(SCr/\kappa, 1)^{-1.209} \times 0.993^{Age} \times 1.018$ [if female] $\times 1.159$ [if black], where SCr is serum creatinine, $\kappa$ is 62 (μmol/L) for females and 80 (μmol/L) for males, $\alpha$ is -0.329 for females and -0.411 for males, min indicates the minimum of $SCr/\kappa$ or 1, and max indicates the maximum of $SCr/\kappa$ or 1.

[c] Antacids include proton pump inhibitors, sodium hydrogen carbonate and compound aluminium hydroxide.

**Table 2 Pharmacokinetic parameter estimates for the final model and Bootstrap results**

| Parameters | Estimates | %RSE[a] | Shrinkage(%) | Bootstrap | |
|---|---|---|---|---|---|
| | | | | Median | 2.5%-97.5% interval [b] |
| OFV | 2597.835 | / | / | 2512.239 | 1876.518-3295.075 |
| Conditional number | 117.84/53.67 | / | / | / | / |
| *Pharmacokinetic parameters* | | | | | |
| $CL_{uMPA}$/F (L/h) | 852 | 7.1 | / | 852.57 | 716.95-989.68 |
| $Q_{uMPA}$/F (L/h)[c] | 859 | 9.0 | / | 843.98 | 728.17-1039.38 |
| Exponent for the effect of BW on $Q_{uMPA}$/F | 2.13 | 15.7 | / | 2.09 | 0.95-3.20 |
| $V_{CuMPA}$/F (L) | 703 | 17.4 | / | 723.20 | 478.32-968.10 |
| $k_a$ (/h) | 1.33 | 9.8 | / | 1.35 | 1.13-1.64 |
| Tlag (h) | 0.445 | 17.6 | / | 0.439 | 0.286-0.628 |
| $k_B$ (/h) [d] | 53.4 | 2.3 | / | 53.33 | 46.29-60.76 |
| Exponent for the effect of ALB on $k_B$ | 1.07 | 11.4 | / | 1.07 | 0.31-1.65 |
| $V_{CuMPAG}$/F (L) | 34.4 | 7.6 | / | 34.23 | 29.28-40.02 |
| $CL_{uMPAG}$/F (L/h) [e] | 6.59 | 4.4 | / | 6.54 | 5.96-7.30 |
| Exponent for the effect of GFR on $CL_{uMPAG}$/F | 0.863 | 11.9 | / | 0.793 | 0.329-1.529 |
| %EHC | 5.13 | 22.8 | / | 5.08 | 2.83-7.77 |
| *Between-subject variability* | | | | | |
| $CL_{uMPA}$/F (%CV) | 50.6 | 17.1 | 3.5 | 49.9 | 41.0-59.8 |
| $Q_{uMPA}$/F (%CV) | 44.7 | 17.9 | 17.4 | 42.8 | 27.3-57.5 |
| $V_{CuMPA}$/F (%CV) | 82.8 | 21.3 | 23.1 | 79.1 | 49.6-108.4 |
| $k_a$ (%CV) | 43.1 | 24.9 | 32.6 | 44.8 | 30.5-64.3 |

*(Continues)*

**Table 2 continued**

| | | | | | |
|---|---|---|---|---|---|
| Tlag (%CV) | 108.2 | 16.8 | 7.9 | 109.7 | 82.9-164.8 |
| $k_B$ (%CV) | 10.0 FIXED | / | / | 10.0 | 10.0-10.0 |
| $V_{CuMPAG}$/F (%CV) | 47.9 | 44.5 | 14.5 | 47.8 | 29.6-65.8 |
| Covariance between $V_{CuMPAG}$/F and $CL_{uMPAG}$/F | 0.0869 | 39.1 | / | 0.0848 | 0.0308-0.1491 |
| $CL_{uMPAG}$/F (%CV) | 31.8 | 11.7 | 2.6 | 32.1 | 22.9-41.7 |
| %EHC (%CV) | 53.9 | 32.9 | 59.7 | 53.6 | 15.9-99.0 |
| *Residual unexplained variability* | | | | | |
| uMPA (%CV) | 47.1 | 3.6 | 5.3 | 47.0 | 41.5-52.2 |
| Covariance between uMPA and tMPA | 0.109 | 14.4 | / | 0.110 | 0.080-0.141 |
| tMPA (%CV) | 45.8 | 3.7 | 5.4 | 45.6 | 40.8-49.5 |
| uMPAG (%CV) | 22.0 | 3.2 | 4.9 | 21.4 | 18.4-24.3 |

MPA, mycophenolic acid; MPAG, 7-O-mycophenolic acid glucuronide; tMPA, total MPA; tMPAG, total MPAG; uMPA, unbound MPA; uMPAG, unbound MPAG; /, not applicable; %CV, percentage coefficient of variation; %EHC, percentage of MPAG recycled into the systemic circulation; %RSE, percentage relative standard error; ALB, serum albumin; BSV, between-subject variability; BW, body weight; $CL_{uMPA}$/F, apparent clearance of uMPA; $CL_{uMPAG}$/F, apparent clearance of uMPAG; GFR, glomerular filtration rate; $k_a$, absorption rate constant; $k_B$, protein binding rate constant; OFV, objective function value; $Q_{uMPA}$/F, intercompartmental clearance of uMPA; RUV, residual unexplained variability; Tlag, lagged absorption time; $V_{CuMPA}$/F, apparent central volume of distribution of uMPA; $V_{CuMPAG}$/F, apparent central volume of distribution of uMPAG

[a] %RSE is estimated as the standard error of the estimate divided by the population estimate multiplied by 100.

[b] Based on 495/500 successful bootstrap runs.

[c] $Q_{uMPA}/F = 859 \times \left[\frac{BW}{70}\right]^{2.13} \quad (L/h)$

[d] $k_B = 53.4 \times \left[\frac{ALB}{40}\right]^{1.07} \quad (/h)$

[e] $CL_{uMPAG}/\mathrm{F} = 6.59 \times \left[\frac{GFR}{80}\right]^{0.863} \quad (L/h)$

The disposition parameter estimates for tMPA and tMPAG are generated by multiplying the unbound concentration based parameters in the original model by the typical unbound fraction at serum albumin concentration of 40 g/L:

$$FU_{MPA} = \frac{1}{1 + k_B} = \frac{1}{1 + 53.4} = 1.84\%$$

$$CL_{tMPA}/F = CL_{uMPA}/F \times FU_{MPA} = 852 \times 1.84\% = 15.68\ (L/h)$$

$$Q_{tMPA}/F = Q_{uMPA}/F \times FU_{MPA} = 859 \times \left[\frac{BW}{70}\right]^{2.13} \times 1.84\% = 15.81 \times \left[\frac{BW}{70}\right]^{2.13} \quad (L/h)$$

$$V_{CtMPA}/F = V_{CuMPA}/F \times FU_{MPA} = 703 \times 1.84\% = 12.94\ (L)$$

$$V_{PtMPA}/F = V_{PuMPA}/F \times FU_{MPA} = 34300 \times 1.84\% = 631.12\ (L)$$

$$CL_{tMPAG}/F = CL_{uMPAG}/F \times FU_{MPAG} = 6.59 \times \left[\frac{GFR}{80}\right]^{0.863} \times 18\% = 1.19 \times \left[\frac{GFR}{80}\right]^{0.863} \quad (L/h)$$

$$V_{CtMPAG}/F = V_{CuMPAG}/F \times FU_{MPAG} = 34.4 \times 18\% = 6.12\ (L)$$

where, $CL_{tMPA}$/F is apparent clearance of tMPA, $CL_{tMPAG}$/F is apparent clearance of tMPAG, $FU_{MPA}$ is unbound fraction of MPA, $FU_{MPAG}$ is unbound fraction of MPAG, $Q_{tMPA}$/F is intercompartmental clearance of tMPA, $V_{CtMPA}$/F is apparent central volume of distribution of tMPA, $V_{CtMPAG}$/F is apparent central volume of distribution of tMPAG and $V_{PtMPA}/F$ is apparent peripheral volume of distribution of tMPA

**Table 3 Previously published population pharmacokinetic analysis of unbound and total mycophenolic acid**

| References | Present study | Okour *et al*, 2018[30] | Colom *et al*, 2018[29] | van Hest *et al*, 2009[28] | de Winter *et al*, 2009[27] |
|---|---|---|---|---|---|
| Number of patients | 58 | 92 | 56 | 88 | 75 |
| Concomitant CNI | CsA | CsA /TAC | CsA /TAC | CsA | CsA /TAC |
| Postoperative time | 3-3084 days | / | 7 days~1 year | 7~148 days | 4~155 days |
| Structure model | MPA: 2 CMT<br>MPAG: 1 CMT | MPA: 1 CMT<br>MPAG: 1 CMT<br>acyl-MPAG :1 CMT | MPA: 2 CMT | MPA: 2 CMT<br>MPAG: 2 CMT | MPA: 2 CMT<br>MPAG: 1 CMT |
| *pharmacokinetic parameter* [a] | | | | | |
| $FU_{MPA}$ (%) | 1.84 [b] | 2.4 | 1.93 [b] | 2.03 [b] | / |
| $CL_{uMPA}/F$ (L/h) | 852 | 1832 | 654 | 866 | 747 |
| $V_{CuMPA}/F$ (L) | 703 | 5630 | 18.3 | 2990 | 189 |
| $Q_{uMPA}/F$ (L/h) | 859 | / | 749 | 1210 | 2010 |
| $V_{PuMPA}/F$ (L) | 34300 FIXED | / | 29100 | 6240 | 34300 |
| *Between-subject variability* | | | | | |
| $CL_{uMPA}/F$ (%CV) | 50.6 | 30.1 | 26.81 | 25 | 97 |
| $V_{CuMPA}/F$ (%CV) | 82.8 | 35.5 | 99.45 | 91 | 116 |
| *Between-occasion variability* | | | | | |
| $CL_{uMPA}/F$ (CV%) | / | / | 40.9 | / | / |
| $V_{CuMPA}/F$ (CV%) | / | / | 137.6 | / | / |
| *Residual unexplained variability* | | | | | |
| uMPA | Exponential: 47.1% | proportional: 40.5% | proportional: 58.3% | log-add: 0.44 | log-add: 0.993 |
| tMPA | Exponential: 45.8% | proportional: 35.8% | proportional: 46.9% | log-add: 0.42 | log-add: 0.52 |

MPA, mycophenolic acid; MPAG, 7-O-mycophenolic acid glucuronide; tMPA, total MPA; uMPA, unbound MPA; /, not applicable or not available; %CV, percentage coefficient of variation; acyl-MPAG, acyl-glucuronide mycophenolic acid; $CL_{uMPA}/F$, apparent clearance of uMPA; CMT, compartment; CNI, calcineurin inhibitor; CsA, ciclosporin; $FU_{MPA}$, unbound fraction of MPA; log-add, log-transformed additive; $Q_{uMPA}/F$, intercompartmental clearance of uMPA; TAC, tacrolimus; $V_{CuMPA}/F$, apparent central volume of distribution of uMPA; $V_{PuMPA}/F$, apparent peripheral volume of distribution of uMPA

[a] Represented as typical reference subjects: 1) body weight 70 kg, 2) serum albumin concentration 40 g/L, 3) glomerular filtration rate 90 mL/min, 4) co-treated with CsA 300 mg per day, 5) total MPAG concentration 0.1 mmol/L.

[b] calculated based on the protein binding rate constant.

## Figure legends

**Figure 1 Schematic representation of the final structural model characterising the linear protein binding and intermittent EHC processes.** In this model, mealtimes are used as an index of gallbladder emptying. This process is assumed to occurrence at the specific time points (mealtimes) with a first-order rate constant and a certain duration. MMF, mycophenolate mofetil; MPA, mycophenolic acid; MPAG, 7-O-mycophenolic acid glucuronide; tMPA, total MPA; tMPAG, total MPAG; uMPA, unbound MPA; uMPAG, unbound MPAG; ALB, serum albumin; BW, body weight; $D_{GB}$, duration of gallbladder emptying; EHC, enterohepatic circulation; %EHC, percentage of MPAG recycled into the systemic circulation; $FU_{MPA}$, unbound fraction of MPA; $FU_{MPAG}$, unbound fraction of MPAG; GFR, glomerular filtration rate; $k_{23}$, transfer rate constant from uMPA central compartment to peripheral compartment; $k_{24}$, elimination rate constant of uMPA; $k_{32}$, transfer rate constant from uMPA peripheral compartment to central compartment; $k_a$, absorption rate constant; $k_B$, protein binding rate constant; $k_{e0}$, elimination rate constant of uMPAG; $k_{GB}$, gallbladder emptying rate constant; $k_{GG}$, transfer rate constant from uMPAG central compartment to gallbladder; Tlag, lagged absorption time; $V_{CuMPA}/F$, apparent central volume of distribution of uMPA; $V_{CuMPAG}/F$, apparent central volume of distribution of uMPAG; $V_{PuMPA}/F$, apparent peripheral volume of distribution of uMPA

**Figure 2 Goodness-of-fit plots of final model for uMPA, tMPA and tMPAG. (A)** population predictions versus observations; **(B)** individual predictions versus observations; **(C)** population predictions versus conditional weighted residuals; **(D)** time after previous dose versus conditional weighted residuals. Red dashed lines and gray-shaded areas represent the locally weighted regression line and 95% confidence interval, respectively. In plots A and B, black solid lines represent the line of unity. In plots C and D, black solid and dashed lines represent the y=0 and y=±1.96 reference lines, respectively. tMPA, total mycophenolic acid (MPA); tMPAG, total 7-O-mycophenolic acid glucuronide; uMPA, unbound MPA

**Figure 3 Prediction-corrected visual predictive check plots of final model for uMPA, tMPA and tMPAG.** Blue dots represent the observed concentrations. Red solid lines represent the median of observations, and the semitransparent red fields represent the simulation-based 95% CIs for the median. The observed 5th and 95th percentiles are presented with red dashed lines, and the simulation-based 95% CIs for corresponding percentiles are shown as semitransparent blue fields. In general, the median and 5th and 95th percentile lines of observations fall inside the area of the corresponding 95% CIs. Additionally, the majority of observed concentrations fall within the 90% prediction interval, which demonstrates that the predicted variability does not exceed the observed variability. CIs, confidence intervals; tMPA, total mycophenolic acid (MPA); tMPAG, total 7-O-mycophenolic acid glucuronide; uMPA, unbound MPA

**Figure 4 Posterior predictive check graphics of final model for uMPA, tMPA and tMPAG.** The histograms represent the distribution of simulations. Black and blue solid lines represent the medians of observations and simulations, respectively. The observed 5th and 95th percentiles are presented by black dashed lines, and the simulated 5th and 95th percentiles are presented by blue dashed lines. The simulated $AUC_{0\text{-}12h}$ values present good consistency with observations. In particular, the 5th percentiles of simulations and observations for uMPA and tMPAG, as well as the medians of simulations and observations for tMPAG, are completely overlapped in the graphics. $AUC_{0\text{-}12h}$, area under the concentration-time curve within 12-h dose-interval; tMPA, total mycophenolic acid (MPA); tMPAG, total 7-O-mycophenolic acid glucuronide; uMPA, unbound MPA

**Figure 5 Model-predicted covariate effects on $AUC_{0\text{-}12h}$ of uMPA, tMPA and tMPAG.** Black squares represent median values and error bars represent 95% confidence intervals of the normalized exposure ratios relative to the typical reference subject (ALB 40 g/L, GFR 90 mL/min) across 2000 simulation replicates. The vertical red dashed lines show an exposure ratio of 1 relative to the reference subject. ALB, serum albumin; $AUC_{0\text{-}12h}$, area under the concentration-time curve within 12-h dose-interval; GFR, glomerular filtration rate; tMPA, total mycophenolic acid (MPA); tMPAG, total 7-O-mycophenolic acid glucuronide; uMPA, unbound MPA

Figure 1

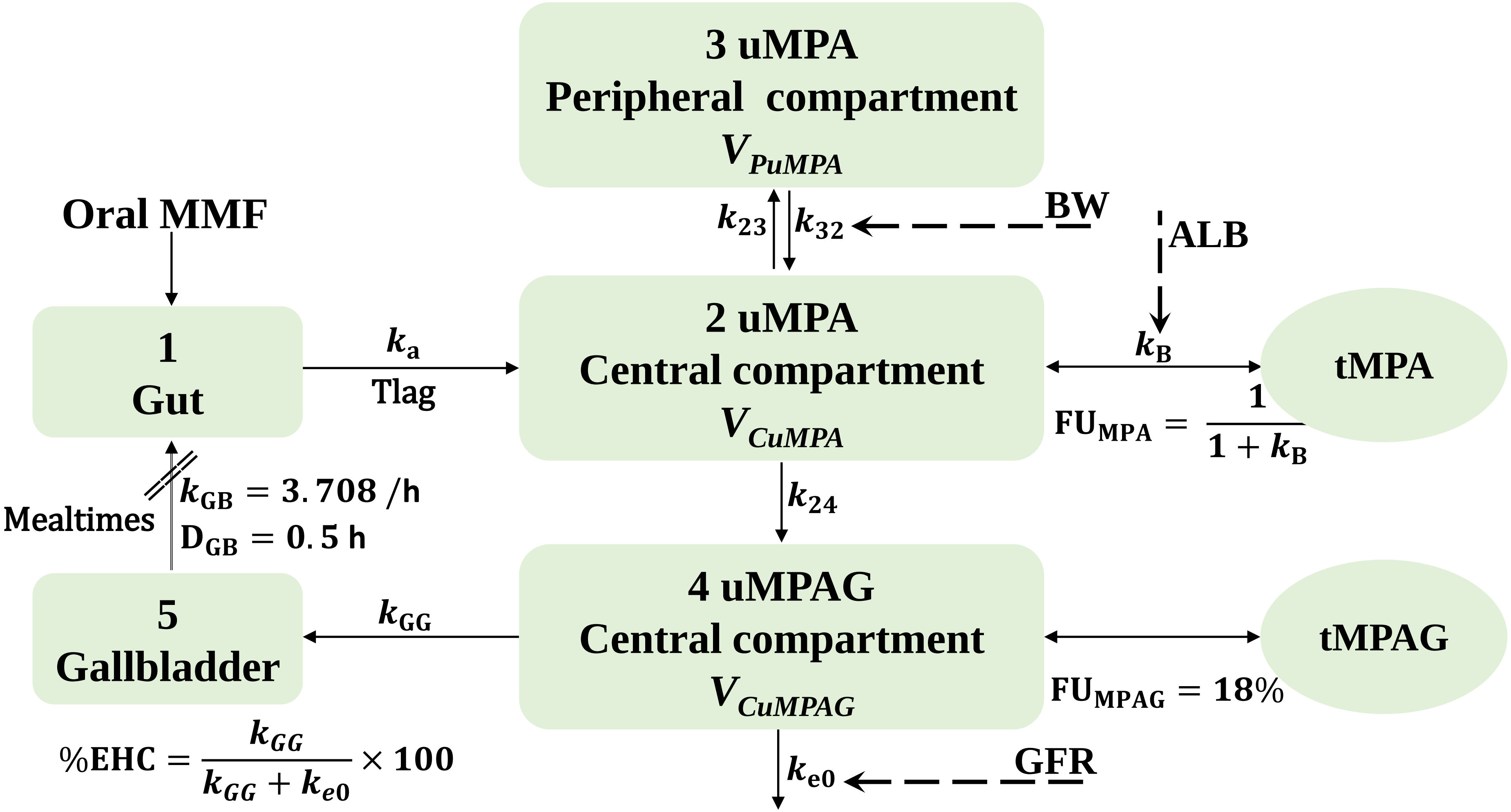

Figure 2

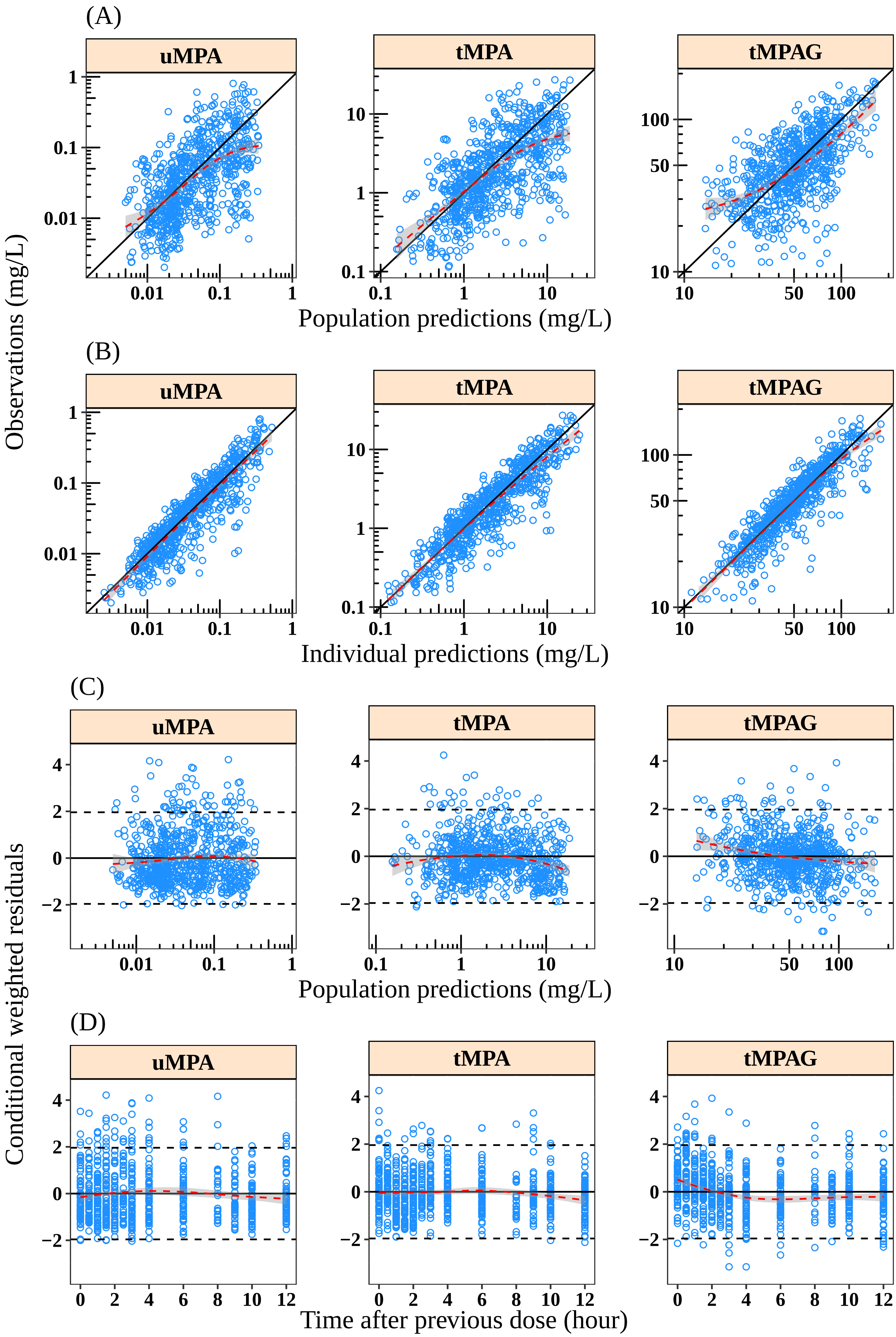

Figure 3

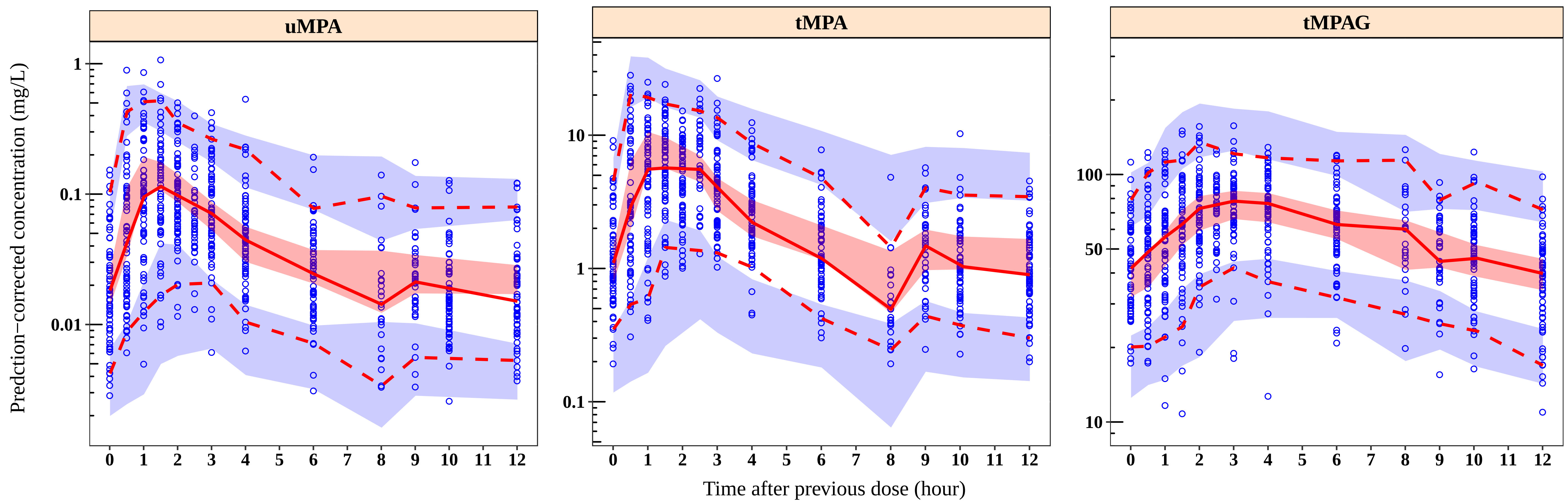

uMPA
tMPA
tMPAG
Predction−corrected concentration (mg/L)
Time after previous dose (hour)
1
0.1
0.01
10
1
0.1
100
50
10
0 1 2 3 4 5 6 7 8 9 10 11 12

Figure 4

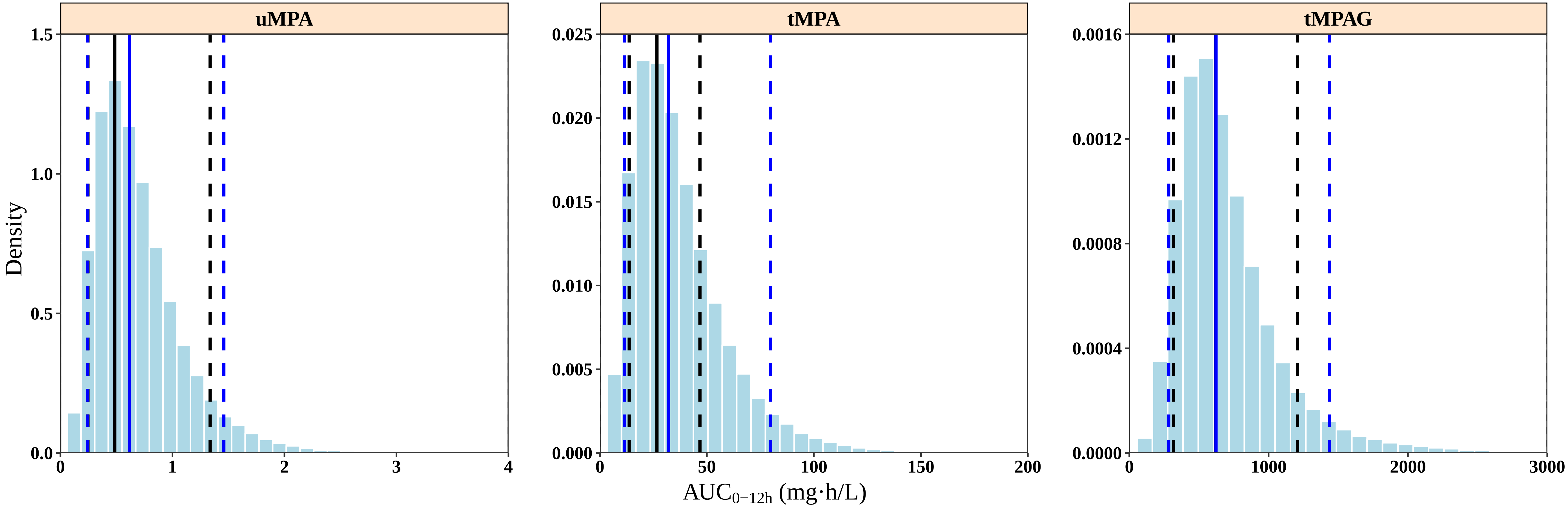

Figure 5

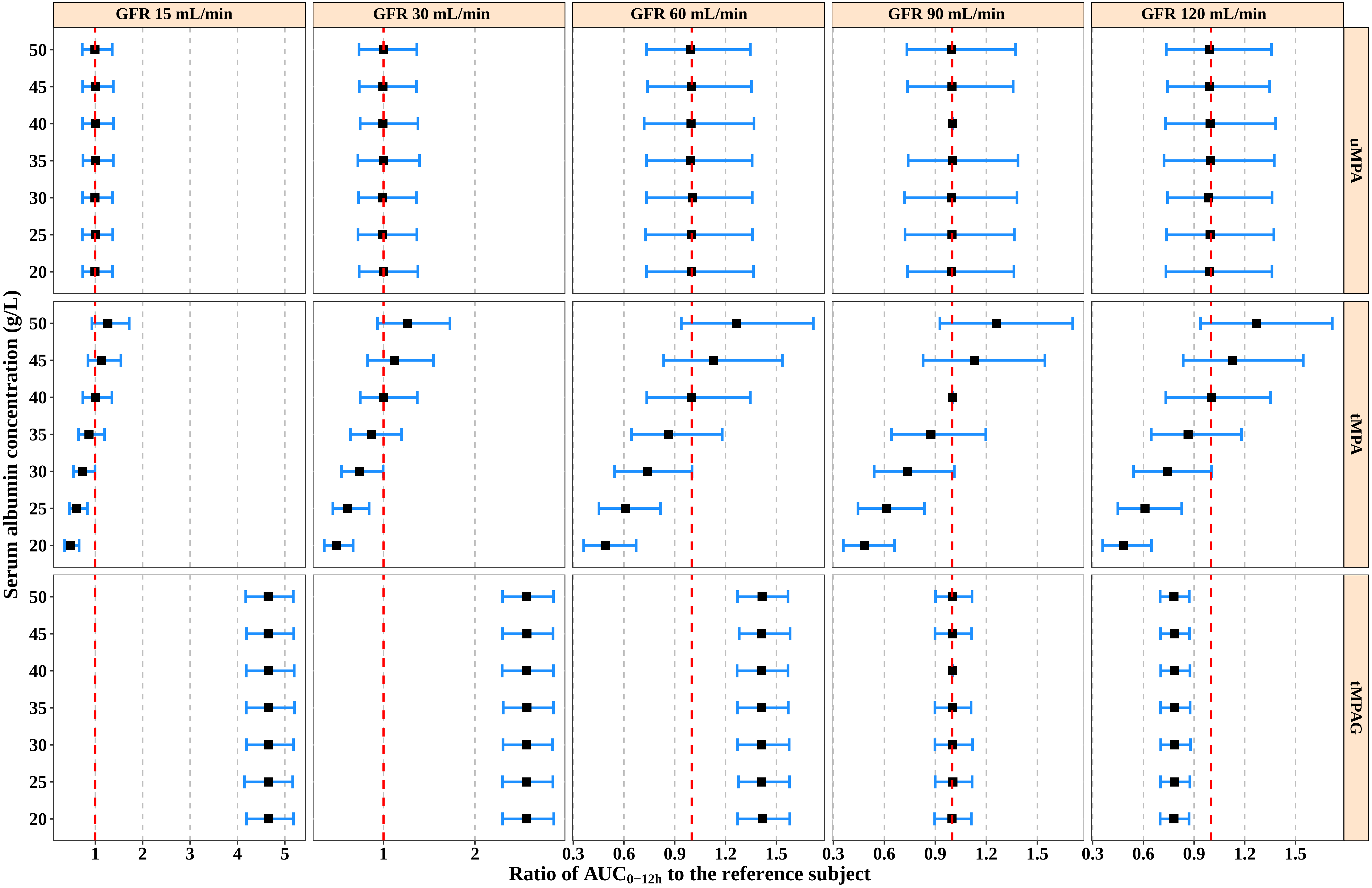

## *Supporting information*

***Table S1***

Key covariate model development steps

***Figure S1***

Model-predicted covariate effect on unbound fraction of MPA. Black squares represent median values and error bars represent 95% confidence intervals of unbound fraction of MPA across 2000 simulation replicates. GFR, glomerular filtration rate; MPA, mycophenolic acid

**Table S1 Key covariate model development steps**

| Model No. | Model description | OFV | ΔOFV | Significance [a] | *df* | Screening results | Comparator |
|---|---|---|---|---|---|---|---|
| 001 | Base model | 2896.614 | / | / | / | / | / |
| ***Forward inclusion step 1*** | | | | | | | |
| 101 | Add BW on $CL_{uMPA}$/F | 2895.068 | -1.546 | No | 1 | / | 001 |
| 102 | Add BW on $Q_{uMPA}$/F | 2870.053 | -26.561 | **Yes** | 1 | / | 001 |
| 103 | Add BW on $V_{CuMPA}$/F | 2897.441 | 0.827 | No | 1 | / | 001 |
| 104 | Add BW on $CL_{uMPAG}$/F | 2895.062 | -1.552 | No | 1 | / | 001 |
| 105 | Add BW on $V_{CuMPAG}$/F | 2898.246 | 1.632 | No | 1 | / | 001 |
| 106 | Add GFR on $k_B$ | 2888.127 | -8.487 | **Yes** | 1 | / | 001 |
| 107 | Add GFR on $CL_{uMPA}$/F | 2900.043 | 3.429 | No | 1 | / | 001 |
| 108 | Add GFR on $Q_{uMPA}$/F | 2898.741 | 2.127 | No | 1 | / | 001 |
| 109 | Add GFR on $CL_{uMPAG}$/F | 2859.631 | -36.983 | **Yes** | 1 | / | 001 |
| 110 | Add ALB on $k_B$ | 2826.643 | -69.971 | **Yes** | 1 | Included | 001 |
| 111 | Add ALB on $V_{CuMPA}$/F | 2885.293 | -11.321 | **Yes** | 1 | / | 001 |
| 112 | Add ALB on $V_{CuMPAG}$/F | 2931.664 | 35.05 | No | 1 | / | 001 |
| 113 | Add ANTAC on $CL_{uMPA}$/F | 2896.066 | -0.548 | No | 1 | / | 001 |
| 114 | Add ANTAC on $k_a$ | 2891.800 | -4.814 | **Yes** | 1 | / | 001 |
| 115 | Add GEND on $CL_{uMPA}$/F | 2901.078 | 4.464 | No | 1 | / | 001 |
| 116 | Add GEND on $Q_{uMPA}$/F | 2875.397 | -21.217 | **Yes** | 1 | / | 001 |
| 117 | Add GEND on $V_{CuMPA}$/F | 2896.181 | -0.433 | No | 1 | / | 001 |
| 118 | Add GEND on $CL_{uMPAG}$/F | 2898.042 | 1.428 | No | 1 | / | 001 |
| 119 | Add GEND on $V_{CuMPAG}$/F | 2897.878 | 1.264 | No | 1 | / | 001 |
| 120 | Add MPAG concentration on $k_B$ | 2898.906 | 2.292 | No | 1 | / | 001 |
| 121 | Add MPAG trough concentration on $k_B$ | 2899.638 | 3.024 | No | 1 | / | 001 |
| ***Forward inclusion step 2*** | | | | | | | |
| 201 | Add BW on $Q_{uMPA}$/F | 2794.131 | -32.512 | **Yes** | 1 | / | 110 |
| 202 | Add GFR on $k_B$ | 2832.015 | 5.372 | No | 1 | / | 110 |
| 203 | Add GFR on $CL_{uMPAG}$/F | 2786.521 | -40.122 | **Yes** | 1 | Included | 110 |
| 204 | Add ALB on $V_{CuMPA}$/F | 2812.156 | -14.487 | **Yes** | 1 | / | 110 |
| 205 | Add ANTAC on $k_a$ | 2819.256 | -7.387 | **Yes** | 1 | / | 110 |
| 206 | Add GEND on $Q_{uMPA}$/F | 2805.960 | -20.683 | **Yes** | 1 | / | 110 |
| ***Forward inclusion step 3*** | | | | | | | |
| 301 | Add BW on $Q_{uMPA}$/F | 2757.313 | -29.208 | **Yes** | 1 | Included | 203 |
| 302 | Add ALB on $V_{CuMPA}$/F | 2766.204 | -20.317 | **Yes** | 1 | / | 203 |
| 303 | Add ANTAC on $k_a$ | 2776.640 | -9.881 | **Yes** | 1 | / | 203 |
| 304 | Add GEND on $Q_{uMPA}$/F | 2767.085 | -19.436 | **Yes** | 1 | / | 203 |
| ***Forward inclusion step 4*** | | | | | | | |
| 401 | Add ALB on $V_{CuMPA}$/F | 2743.568 | -13.745 | **Yes** | 1 | Not included [b] | 301 |
| 402 | Add ANTAC on $k_a$ | 2750.300 | -7.013 | **Yes** | 1 | / | 301 |
| 403 | Add GEND on $Q_{uMPA}$/F | 2749.149 | -8.164 | **Yes** | 1 | Included | 301 |
| ***Forward inclusion step 5*** | | | | | | | |
| 501 | Add ALB on $V_{CuMPA}$/F | 2732.134 | -17.015 | **Yes** | 1 | Not included [b] | 401 |
| 502 | Add ANTAC on $k_a$ | 2739.109 | -10.04 | **Yes** | 1 | Included | 401 |
| ***Forward inclusion step 6*** | | | | | | | |
| 601 | Add ALB on $V_{CuMPA}$/F | 2725.397 | -13.712 | **Yes** | 1 | Not included [b] | 502 |
| ***Backward elimination step 1*** | | | | | | | |
| 1001 | Eliminate ALB from $k_B$ | 2817.487 | 78.378 | **Yes** | 1 | Retained | 502 |
| 1002 | Eliminate GFR from $CL_{uMPAG}$/F | 2778.187 | 39.078 | **Yes** | 1 | Retained | 502 |
| 1003 | Eliminate BW from $Q_{uMPA}$/F | 2758.365 | 19.256 | **Yes** | 1 | Retained | 502 |
| 1004 | Eliminate GEND from $Q_{uMPA}$/F | 2750.300 | 11.191 | **Yes** | 1 | Retained | 502 |
| 1005 | Eliminate ANTAC from $k_a$ | 2749.149 | 10.04 | **No** | 1 | Eliminated | 502 |
| ***Backward elimination step 2*** | | | | | | | |
| 1011 | Eliminate ALB from $k_B$ | 2831.266 | 82.117 | **Yes** | 1 | Retained | 1005 |
| 1012 | Eliminate GFR from $CL_{uMPAG}$/F | 2786.818 | 37.699 | **Yes** | 1 | Retained | 1005 |
| 1013 | Eliminate BW from $Q_{uMPA}$/F | 2767.085 | 17.936 | **Yes** | 1 | Retained | 1005 |
| 1014 | Eliminate GEND from $Q_{uMPA}$/F | 2757.313 | 8.164 | **No** | 1 | Eliminated | 1005 |
| ***Backward elimination step 3*** | | | | | | | |
| 1021 | Eliminate ALB from $k_B$ | 2831.921 | 74.608 | **Yes** | 1 | Retained | 1014 |
| 1022 | Eliminate GFR from $CL_{uMPAG}$/F | 2794.131 | 36.818 | **Yes** | 1 | Retained | 1014 |
| 1023 | Eliminate BW from $Q_{uMPA}$/F | 2786.521 | 29.208 | **Yes** | 1 | Retained | 1014 |

ALB, serum albumin concentration; ANTAC, antacid; BW, body weight; $CL_{uMPA}$/F, apparent clearance of uMPA; $CL_{uMPAG}$/F, apparent clearance of uMPAG; *df*, degree of freedom; GEND, gender; GFR, glomerular filtration rate; $k_a$, absorption rate constant; $k_B$, protein binding rate constant; MPAG, 7-O-mycophenolic acid glucuronide; OFV, objective function value; $Q_{uMPA}$/F, intercompartmental clearance of uMPA; uMPA, unbound mycophenolic acid; uMPAG, unbound MPAG; $V_{CuMPA}$/F, apparent central volume of distribution of uMPA; $V_{CuMPAG}$/F, apparent central volume of distribution of uMPAG; ΔOFV, change in OFV

[a] Significance levels are set at a decrease in OFV of >3.84 ($\chi^2$, *df* =1, *p* <0.05) during the forward inclusion step and an increase in OFV of >10.83 ($\chi^2$, *df* =1, *p* <0.001) during the backward elimination step, respectively.

[b] Addition of ALB on $V_{CuMPA}$ results in a significant decrease in OFV. However, it also leads to an increase in the between-subject variability (BSV) on $V_{CuMPA}$ of over 60%. Thus, ALB is not included as a significant covariate.

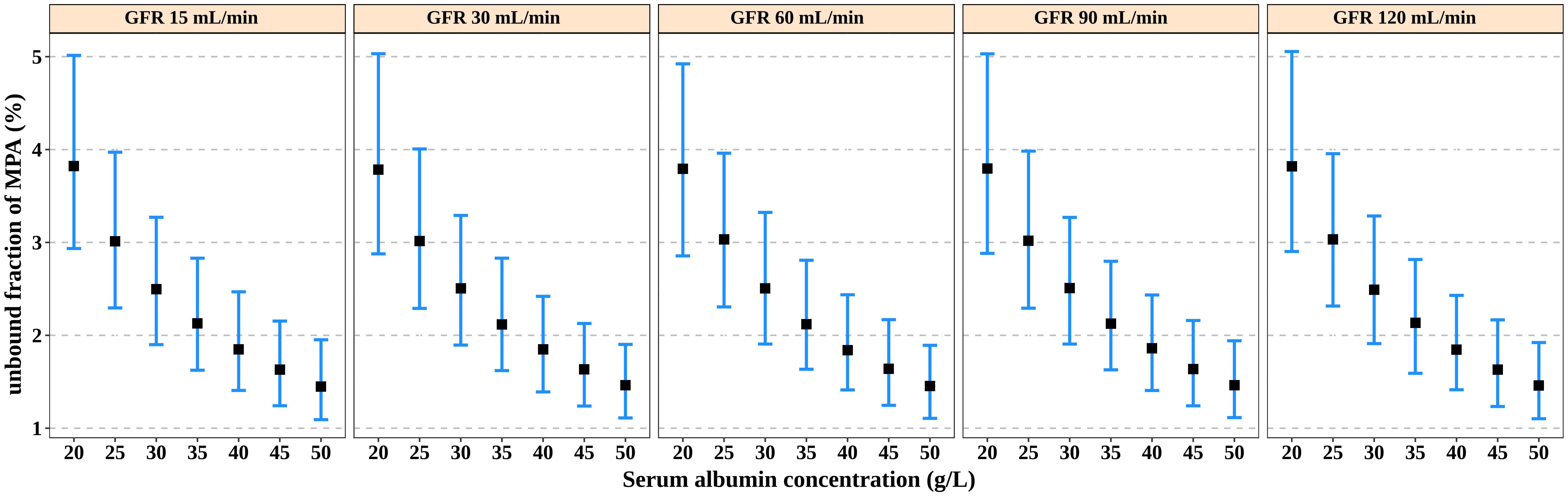

***Figure S1*** Model-predicted covariate effect on unbound fraction of MPA. Black squares represent median values and error bars represent 95% confidence intervals of unbound fraction of MPA across 2000 simulation replicates. GFR, glomerular filtration rate; MPA, mycophenolic acid